\documentclass{ws-procs9x6}

\begin{document}

\title{Inclusive Jet $A_{LL}$ Measurements at STAR}

\author{D. Staszak (for the STAR Collaboration)}

\address{University of California, Los Angeles \\
430 Portola Plaza, Box 951547 \\ 
Los Angeles, CA 90095-1547, USA\\ 
E-mail: staszak@physics.ucla.edu}

\maketitle

\abstracts{
We report STAR's preliminary measurement of the inclusive jet
longitudinal spin asymmetry $A_{LL}$ using data from the RHIC 2006 run.
The 2006 data set was taken with
200 GeV polarized proton-proton collisions 
and represents 4.7 $pb^{-1}$ of data.
Typical beam polarizations were $\sim 55-60\%$.
The data are compared with theoretical
calculations of $A_{LL}$ based on various GRSV models of the polarized
parton distribution functions in the nucleon.  Previous STAR inclusive jet
$A_{LL}$ and cross  section measurements are also presented, as is a
discussion of constraints these data place on the allowed theoretical models.}

\section{Introduction}
The Relativistic Heavy Ion Collider (RHIC) is unique among the world's high energy
physics laboratories by being the first and only accelerator complex to collide
beams of polarized protons.
The RHIC Spin Program\cite{bunce} aims to use polarized p-p collisions
to measure the gluon spin contribution, $\Delta G$, to the proton's spin.  
As measured by polarized Deep Inelastic Scattering (DIS) over three decades, 
the quark and anti-quark spin contributions to the spin of the proton
are known to be only $\sim 20-30\%$\cite{dis}.  The deficit must be
made up from the gluon spin and/or the quark and gluon orbital angular momentum
contributions, all of which are relatively unconstrained by data.

The $\Delta G$ contribution is accessed at RHIC by measuring the double longitudinal spin asymmetry 
$A_{LL} = \Delta\sigma / \sigma = (\sigma^{++} - \sigma^{+-}) / (\sigma^{++} + \sigma^{+-})$,
where ++ and +- refer to the helicity states of the two incoming
proton beams.  $A_{LL}$ can be measured at RHIC in a variety of processes\cite{bunce}.
The results presented here will be from the inclusive jet channel at 
the Solenoidal Tracker At RHIC (STAR\cite{star}) experiment only.
The inclusive jet channel has the advantages of a relatively large cross section and
little sensitivity to fragmentation functions.  Additionally, STAR's large tracking and 
calorimeter coverage ideally suits full jet reconstruction.

\section{Experimental Setup}

Jets are reconstructed at STAR with a midpoint cone algorithm\cite{Cone} using 
a cone radius of 0.4(0.7) in 2005(6) and
data from the Barrel and Endcap Electro-Magnetic Calorimeters (BEMC, EEMC) 
and the Time Projection Chamber (TPC).  
The BEMC and EEMC measure the neutral energy deposited by each event in the 
full eta range of $-1 < \eta < 2$. 
The BEMC consists of 4800 towers of alternating layers of lead and scintillator each 
$0.05 \times 0.05$ in $\eta$, $\phi$; the EEMC consists of 720 towers with 
varying coverage $(0.05-0.1)\times0.1$ in  $\eta$, $\phi$.
Charged track momenta are measured in the 0.5 T STAR solenoidal magnetic field
by the TPC for $-1.3 \lesssim \eta \lesssim 1.3$.
For triggering and
beam luminosity monitoring, Beam Beam Counters (BBCs) are used in the 
range $3.4 < |\eta| < 5$ on both sides of the collision point.
Each BBC consists of two concentric arrays of hexagonal scintillator tiles,
and a coincident signal between any tiles from the East and West 
detectors forms STAR's Minimum Bias (MB) trigger.

The data presented here were taken during two extended p-p runs in
2005 and 2006.
In order to increase the relative population of the higher $p_T$ jets in our samples,
we heavily prescale events that pass the MB trigger alone and add additional BEMC
trigger conditions.  
The 2005 results shown utilized High Tower (HT) triggers and Jet Patch (JP)
triggers, while 2006 results shown are from JP only.
The low(high) HT trigger in 2005 required each accepted event to 
have a single BEMC tower with transverse energy $E_T \geq 2.6(3.5)$ GeV.  
The JP trigger in 2005 required $E_T \geq 4.5(6.5)$ GeV within a BEMC region 
$1\times1$ in $\eta$, $\phi$, and in 2006 required $E_T \geq 7.8(8.3)$ GeV
within the JP per accepted event.

\section{$A_{LL}$ Results and Comparison to Theory}

Previous p-p inclusive jet results have been published  from
STAR\cite{jetspaper} in which good agreement within systematics was
found between the measured cross section for STAR data using two
trigger arrangements and an NLO pQCD evaluation\cite{jager}.
The agreement confirms the applicability of the NLO pQCD framework 
to STAR's energy regime and supports interpretation of $A_{LL}$ in this framework.
Partonic subprocesses including $gg \rightarrow gg$, 
$qg \rightarrow qg$, and $qq \rightarrow qq$ all contribute in a $p_T$ 
dependent way to $A_{LL}$ at leading order. 
Subprocesses including a gluon, $gg$ and $qg$, dominate in STAR's $p_T$ and
kinematic range\cite{jager}$^{,}$\cite{bunce}.

\begin{figure}[ht]
\epsfxsize=10cm   
\centerline{\epsfxsize=3.7in\epsfbox{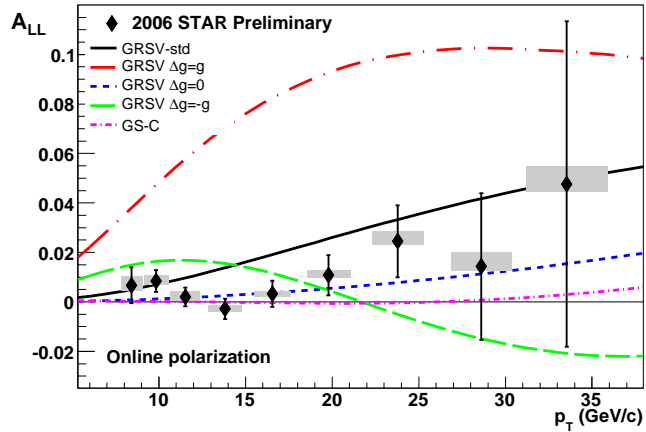}}   
\caption{$A_{LL}$ as a function of corrected jet transverse momentum 
for inclusive jet production at $\sqrt{s}$ = 200 GeV.
The error bars represent statistical uncertainties only. The grey bands represent the
systematic uncertainties, excluding an overall uncertainty for
beam polarization.  Data are shown compared to NLO
polarized gluon models from GRSV$^7$ and GS$^{10}$.}
\label{ALL}
\end{figure}

Experimentally, $A_{LL}$ is measured as:

\begin{equation}
A_{LL} = \frac{1}{P_1 P_2} \frac{N^{++} - R N^{+-}}
{N^{++} + R N^{+-}}, 
\; \; \; R = \frac{\mathcal{L}^{++}}{\mathcal{L}^{+-}}
\end{equation}

\noindent
where $N^{(++),(+-)}$ are the jet yields for the aligned and anti-aligned
beam helicity states, R is the ratio of luminosities, and $P_{1,2}$ are
the beam polarizations measured with RHIC polarimeters\cite{polar}.  Figure \ref{ALL} shows
$A_{LL}$ from 2006 JP trigger data representing
4.7 $pb^{-1}$ of data after all cuts.  
Beam background not associated with the hard scattering occasionally resulted 
in jet-like signals in the detectors exhibiting an excess of neutral energy.
We remove them by placing a conservative
upper limit on the fraction of the total energy in a jet coming from our calorimeters;
less than 0.80 in 2005 and less than 0.85 in 2006 after shielding was added
around the beampipe upstream of STAR's detectors.  
BBC coincidence timing information is used in 2005(6) to keep events within
uniform TPC tracking efficiency for both yield and relative luminosity measurements.
A cut is also applied to the jet thrust axis in both samples to keep jets within
uniform calorimeter acceptance, $0.2 < \eta < 0.8$ in 2005 and 
$-0.7 < \eta < 0.9$ in 2006.  In 2006 the BEMC was fully instrumented and
both BEMC and EEMC towers were included in jet finding/reconstruction.

\begin{figure}[ht]
\epsfxsize=10cm   
\centerline{\epsfxsize=3.6in\epsfbox{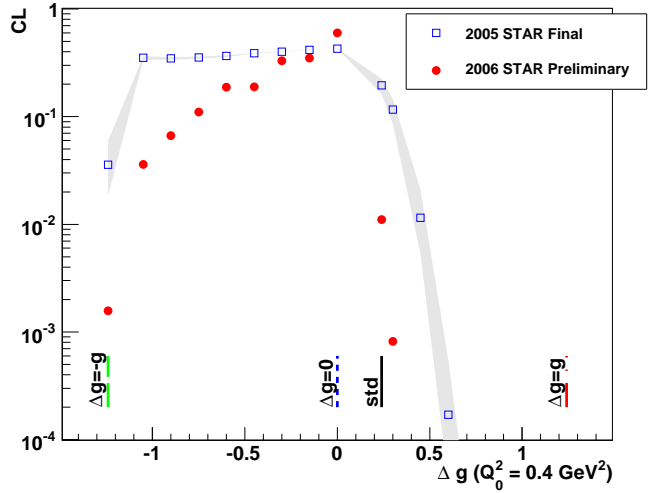}}   
\caption{
2005/6 Confidence Levels are shown for GRSV models ranging from 
$\Delta g = -g$ to $g$. 
The quantity $\Delta g$ represents the integral of x from $0\rightarrow1$
of $\Delta g(x)$ for that model at the input scale $Q^2$=0.4 $GeV^2$.
The shaded band around 2005 points is the $9.4\%$ scale uncertainty 
from beam polarization.}
\label{CL}
\end{figure}

The leading systematic uncertainties on our $A_{LL}$ data points come from biases
introduced by jet reconstruction and triggering.  We determine them
by combining Pythia\cite{pythia} generated events with polarized parton
weights from GRSV\cite{grsv} $\Delta G$ models and a full Geant\cite{geant} STAR
detector and trigger simulation.  
Shifts have been applied to the $p_T$ values of the points in Fig. 1
in order to correct for jet energy scale and resolution effects.
A conservative uncertainty is applied
to each $p_T$ point individually by taking the maximal deviation of all allowed GRSV
$A_{LL}$ theory curves from simulated $A_{LL}$ curves corrected to our detector+trigger.
Other systematic effects studied include 
the possible contribution caused by residual transverse beam polarization components,
the uncertainty on relative luminosities, and the uncertainties
introduced from beam backgrounds.  Single spin asymmetries
are also measured and found to be consistent with zero as expected.

The curves in Fig. \ref{ALL} are NLO pQCD calculations of the inclusive 
jet $A_{LL}$ from two separate theory groups, GRSV\cite{grsv} and GS\cite{gs}.
Within the GRSV framework, GRSV-STD is the best fit to polarized inclusive DIS data
as of 2001.  GRSV $\Delta g = (-g)g$ are gluon polarization models with all the gluons 
predicted to be (anti-)aligned with the proton helicity at the initial scale
$Q^2 = 0.4$ $GeV^2$.  GRSV $\Delta g = 0$ represents
no net gluon polarization within the proton at the initial scale. GS-C contains a node at 
$x\sim 0.1$ for $Q^2 = 4$ $GeV^2$.  STAR's kinematic 
x-range is $0.03<x<0.3$, which accounts for $\sim$ 50\% of the total $\Delta G$
integral for GRSV-STD.

Figure \ref{CL} shows the confidence level (CL) of the comparison of the 2005/6
data and theory for each of the GRSV parameterizations.
Stratmann and Vogelsang provided STAR with additional $\Delta G$ 
models lying between the 4 major GRSV models shown on Fig. \ref{ALL}.  Within 
the GRSV framework, STD can be excluded based on STAR data
with a 99\% CL and gluon polarizations $\Delta g < -0.7$ can be 
excluded with a 90\% CL.  The 1$\sigma$ limits for the best fit
GRSV-STD are $\Delta g =$ -0.45 to 0.7.  Experimental systematic and statistical uncertainties 
were accounted for in the CL calculation. 

\section{Conclusion}

Preliminary 2006 inclusive jet $A_{LL}$ results using $\sqrt{s} = 200$ GeV polarized 
proton-proton collisions with jet $p_T$ up to 35 $GeV/c$ were shown.
4.7 $pb^{-1}$ of data were used with typical beam polarizations ranging $\sim 55-60\%$.
The data provide significant new constraints as illustrated by a CL analysis using GRSV
polarized gluon parameterizations.
The central value of the best fit to DIS data, GRSV-STD, can 
be excluded at 99\% CL, and GRSV models predicting $\Delta g < -0.7$ are excluded
at 90\% CL.

\end{document}